\documentclass[preprint,showpacs,preprintnumbers,amsmath,amssymb]{revtex4}

\usepackage{graphicx}
\usepackage{wasysym}
\usepackage{dcolumn}
\usepackage{bm}

\begin{document}


\title{Solving the stellar $^{62}$Ni problem with AMS}

\date{\today}

\author{I. Dillmann}
 \altaffiliation[previous address: ]{Institut f\"ur Kernphysik, Forschungszentrum Karls\-ruhe, Postfach 3640, D-76021 Karls\-ruhe, Germany}
\author{T. Faestermann}
\author{G. Korschinek}
\author{J. Lachner}
\author{M. Maiti}
	\altaffiliation[now at ]{Saha Institute of Nuclear Physics, Chemical Sciences Division, 1/AF Bidhannagar, Kolkata, India}
\author{M. Poutivtsev}
\author{G. Rugel}
 \affiliation{Physik-Department E12 \& E15 and Excellence Cluster Universe, Research Area G, Technische Universit\"at M\"unchen, D-85748 Garching, Germany}
 
\author{S. Walter}
\author{F. K\"appeler}
 \affiliation{Institut f\"ur Kernphysik, Forschungszentrum Karlsruhe, Postfach 3640, D-76021 Karlsruhe, Germany}

\author{M. Erhard}
\author{A.R. Junghans}
\author{C. Nair}
\author{R. Schwengner}
\author{A. Wagner}
 \affiliation{Institut f\"ur Strahlenphysik, Abt. Kernphysik, Forschungszentrum Dresden-Rossendorf, Germany}
 
\begin{abstract}
An accurate knowledge of the neutron capture cross sections of $^{62,63}$Ni is crucial since both isotopes take key positions which affect the whole reaction flow in the weak $s$ process up to $A$$\approx$90. No experimental value for the $^{63}$Ni$(n,\gamma)$ cross section exists so far, and until recently the experimental values for $^{62}$Ni$(n,\gamma)$ at stellar temperatures ($kT$=30 keV) ranged between 12 and 37 mb. This latter discrepancy could now be solved by two activations with following AMS using the GAMS setup at the Munich tandem accelerator which are also in perfect agreement with a recent time-of-flight measurement. The resulting (preliminary) Maxwellian cross section at $kT$=30~keV was determined to be $<$$\sigma$$>$$_{30keV}$= 23.4 $\pm$ 4.6~mb. 

Additionally, we have measured the $^{64}$Ni$(\gamma,n)$$^{63}$Ni cross section close to thres\-hold. Photoactivations at 13.5~MeV, 11.4~MeV and 10.3~MeV were carried out with the ELBE accelerator at Forschungszentrum Dresden-Rossendorf. A first AMS measurement of the sample activated at 13.5~MeV revealed a cross section smaller by more than a factor of 2 compared to NON-SMOKER predictions.
\end{abstract}


\maketitle

\section{Introduction}
\begin{figure}[!htb]
\includegraphics[scale=0.75]{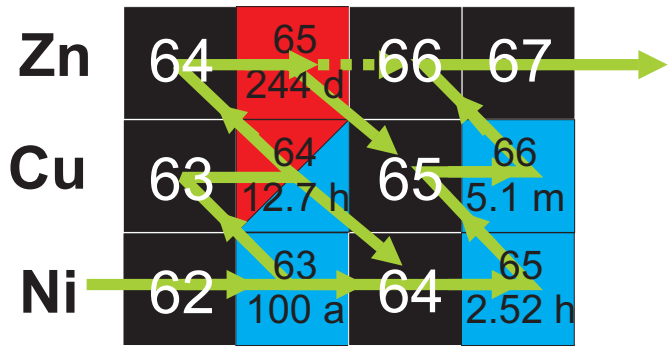}
\caption{Branching of the weak $s$ process at $^{63}$Ni.}\label{branch}
\end{figure}

The nucleosynthesis of elements heavier than iron can be almost completely ascribed to the $s$ process ("slow neutron capture process") and the $r$ process ("rapid neutron capture process") \cite{bbfh57}. The $s$ process can be further divided into a "weak" component (responsible for nuclei up to $A$$\approx$90) and a "main" component (for 90$<$$A$$<$209), which occur in different astrophysical scenarios at different temperatures and with different neutron exposures. The weak $s$-process occurs during core He and shell C burning in massive stars. Among the nuclei involved, the long-lived isotopes $^{63}$Ni (t$_{1/2}$= 100 yr), $^{79}$Se (t$_{1/2}$ $\approx$480000 yr), and $^{83}$Kr (t$_{1/2}$= 10.76 yr) assume key positions, because their stellar $\beta$$^-$-decay rate becomes comparable to the neutron capture rate ($\lambda_\beta$$\approx$$\lambda_n$). The resulting competition leads to branchings in the $s$-process nucleosynthesis path (see Fig.~\ref{branch}).

In the case of $^{63}$Ni the split of the reaction path causes part of the reaction flow to bypass $^{64}$Zn, whereas at $^{66}$Zn both flows merge back. The reaction flow to the higher mass region depends not only on the abundances of the seed nuclei but on their stellar neutron capture cross sections as well. Generally, the (resonant capture part of the) cross section can be determined directly by measuring the prompt $\gamma$-rays associated with neutron capture events in time-of-flight (TOF) experiments. A method to include also the direct capture (DC) component of the capture process is to measure the activity of the product nuclei - if unstable - after the irradiation, applying the so called activation method. The determination of neutron cross sections by means of the activation technique represents an important complement to measurements using the TOF method since this independent approach implies different systematic uncertainties. In combination with accelerator mass spectrometry (AMS) the activation technique can be extended to hitherto inaccessible cases, e.g. to reactions producing very long-lived nuclei with very weak or completely missing  $\gamma$-transitions. The application of AMS counting in stellar neutron reactions has the further advantage of being independent of uncertain $\gamma$-ray intensities.

\section{Cross section measurements}\label{exp}
\subsection{Neutron activations}
The present activations \cite{SW08} and those by Nassar et al. \cite{HN05} (see Table~\ref{tab:act}) were carried out at the (now closed) 3.7 MV Van de Graaff accelerator at Forschungszentrum Karlsruhe using enriched $^{62}$Ni samples. Neutrons were produced with the $^7$Li($p,n$)$^7$Be source by bombarding 30 $\mu$m thick layers of metallic Li on a water-cooled Cu backing with protons of 1912 keV, 30 keV above the reaction threshold. The angle-integrated neutron spectrum imitates almost perfectly a Maxwell-Boltzmann distribution for $kT$ = 25.0$\pm$0.5~keV with a maximum neutron energy of 108~keV \cite{raty88}. 
At this proton energy the neutron flux is kinematically collimated in a forward cone with 120$^\circ$ opening angle. Neutron scattering through the Cu backing is negligible since the transmission is about 98\% in the energy range of interest. To ensure homogeneous illumination of the entire surface, the proton beam with a DC current of $\approx$100~$\mu$A was wobbled across the Li target. The mean neutron f\-lux over the period of the activations was
$\approx$1--2$\times$10$^9$ s$^{-1}$ at the position of the samples, which were placed in close geometry to the Li target. A $^6$Li-glass monitor at 1~m distance from the neutron target was used to record the time-dependence of the neutron yield in intervals of 1~min as the Li target degrades during the irradiation. In this way the proper correction of the number of nuclei which decayed during the activation can be attained. This correction is negligible for very
long half-lives but becomes important for comparably short-lived isotopes like $^{198}$Au, since gold is used as the reference cross section for the neutron flux determination.

\subsection{AMS measurements}
The $^{63}$Ni/$^{62}$Ni ratio of our two independently irradiated $^{62}$Ni samples \cite{SW08} was measured with the gas-filled analyzing magnet system (GAMS) at the 14~MV MP tandem accelerator in Munich. The AMS setup for $^{63}$Ni consists of a cesium sputter ion source dedicated exclusively to this isotope with an isobaric $^{63}$Cu background smaller by 2 orders of magnitude compared to our standard ion sources \cite{RFK00,RAC04}. The sample material was metallic $^{62}$Ni powder from different batches with enrichments of 95\% and 97.3\%, respectively. The samples were pressed into ultrapure graphite cathodes without further chemical processing. The nickel was extracted as Ni$^-$ from the ion source and stripped to positive charge states by a thin carbon foil (4 $\mu$g/cm$^2$) at the terminal of the tandem accelerator. For our measurements we chose the 12$^+$ and 13$^+$ charge states and terminal voltages between 12.4 and 12.7~MV. The GAMS detection system consists of a 135$^\circ$ gas-filled magnet and a segmented Frisch-grid ionization chamber, where the anode is divided along the flight path into 5 energy loss signals $\Delta$E1 to $\Delta$E5. The field of the GAMS magnet was adjusted to suppress the remaining $^{63}$Cu background as much as possible without hampering the $^{63}$Ni measurement. This combination allows the determination of $^{63}$Ni/Ni ratios as low as 2$\times$10$^{-14}$, with a total isobaric suppression of several times 10$^9$ \cite{RFK00,RAC04}.

\begin{figure}[!htb]
\includegraphics[scale=1]{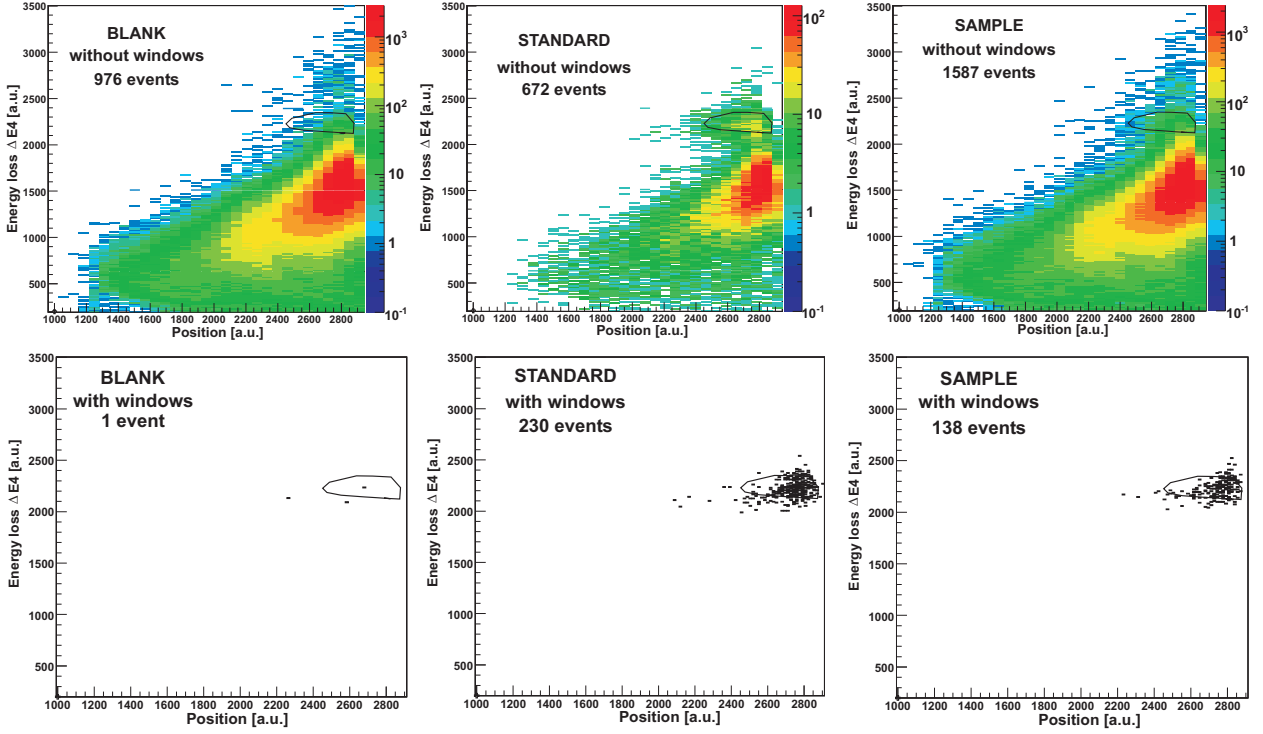}
\caption{Energy loss in detector segment 4 vs. position of blank ($\frac{63Ni}{62Ni}$=3$\times$10$^{-13}$; $\bar{I}$=90~nA; t$_{meas}$=1666~s), standard ($\frac{63Ni}{62Ni}$=3.5$\times$10$^{-9}$; $\bar{I}$=2~nA; t$_{meas}$=1758~s), and activated sample ($\frac{63Ni}{62Ni}$=5.1$\times$10$^{-11}$; $\bar{I}$=79~nA; t$_{meas}$=1578~s). Top: without windows; Bottom: with windows on all other parameters (suppression of $^{63}$Cu background).}\label{ams}
\end{figure}

The activated samples, blank samples, and standards were measured alternately several times under the same conditions (see Fig.~\ref{ams}). 

\section{Results and Discussion}
\subsection{$^{62}$Ni$(n,\gamma)$$^{63}$Ni}
The AMS measurements yielded isotope ratios of $\frac{63Ni}{62Ni}$= (6.0$\pm$1.7)$\times$ 10$^{-11}$ for sample ni-62a and (5.1$\pm$1.3)$\times$ 10$^{-11}$ for ni-62b (Table~\ref{tab:act}). The experimental cross section can be directly obtained from this isotopic ratio via $\sigma_{exp}$= $\frac{63Ni}{62Ni}$$\cdot$ $\frac{1}{\Phi_{tot}}$. Transformation of this weighted experimental cross section $\sigma_{exp}$= 25.2~mb in the quasi-Maxwellian neutron spectrum into a true Maxwellian (for a detailed description of the transformation, see \cite{raty88}) yields a (preliminary) cross section at $kT$=30~keV of $<$$\sigma$$>$$_{30keV}$=23.4$\pm$4.6~mb \cite{SW08}, in perfect agreement with the previous AMS measurement from \cite{HN05} ($<$$\sigma$$>$$_{30keV}$=26.1$\pm$2.6~mb). The counts of mass 63 from the blank sample (see Fig.~\ref{ams}) correspond to an isotope ratio $\frac{63Ni}{62Ni}$=3.0$\times$10$^{-13}$. The origin of these counts is due to the Cu impurity in the blank, but since the number of these background counts is much smaller than the uncertainties in the measurements we did not correct for that.

\begin{table}[!htb]
\caption{Irradiation times "$t_a$", total neutron exposures "$\Phi_{tot}$", resulting isotope ratios $\frac{63Ni}{62Ni}$, and experimental neutron cross section $\sigma_{exp}$ compared to the activation of Ref.~\cite{HN05} which was determined at Argonne National Laboratory with the gas-filled magnet. Quoted errors are from the AMS measurement only. \label{tab:act}}
\renewcommand{\arraystretch}{1.0} 
\begin{ruledtabular}
\begin{tabular}{ccccc}
Activation & $t_a$ [d] & $\Phi_{tot}$ [n] & Ratio $\frac{63Ni}{62Ni}$ & $\sigma_{exp}$ [mb] \\
\hline 
ni-62a & 10.2  	& 1.735$\times$10$^{15}$ & 	(6.0$\pm$1.7)$\times$10$^{-11}$ & 34.8$\pm$9.8 \\
ni-62b & 10.3 	& 2.312$\times$10$^{15}$ & 	(5.1$\pm$1.3)$\times$10$^{-11}$ & 22.6$\pm$5.6 \\
\hline
error weighted & 	& 	&														& 25.2$\pm$4.9 \\
\hline
\hline
ni-62 \cite{HN05} & 17.4  & 3.680$\times$10$^{15}$ &  8.78$\times$10$^{-11}$ & 23.9$\pm$2.4 \\
\end{tabular}
\end{ruledtabular}
\end{table}

\subsection{$^{64}$Ni$(\gamma,n)$$^{63}$Ni}
For the photodisintegration measurements of $^{64}$Ni at the ELBE accelerator at Forschungszentrum Dresden-Rossendorf we activated 2 metallic samples of $^{64}$Ni (enrichment 99.63\%) at 11.5~MeV and 13.4~MeV, and another sample with 89.8\% enrichment at 10.3~MeV. Up to now only one of these samples ($E_\gamma$=13.4~MeV) was measured for a short time with AMS \cite{SW08}. Unfortunately the $^{64}$Ni samples with 99.63\% enrichment had a factor of 10 higher $^{63}$Cu background than the $^{62}$Ni samples. The measured isotopic ratio of $\frac{63Ni}{64Ni}$=(1.1$^{+0.5}_{-0.4}$)$\times$10$^{-12}$ is governed by this large background corrections. The (preliminary) cross section is lower by a factor of two compared to the prediction of NON-SMOKER \cite{NS00}. However, it has to be emphasized here, that this measurement suffered from rather large statistical uncertainties due to the $^{63}$Cu background. A more accurate value is expected after the measurements of all three samples are completed.

\subsection{Astrophysical discussion}
An accurate knowledge of the stellar neutron capture cross sections of $^{62,63}$Ni is required since these two cross sections affect the entire weak $s$-process flow towards heavier nuclei. Especially the $^{62}$Ni$(n,\gamma)$$^{63}$Ni cross section was found to act like a bottleneck due to its previously recommended low cross section of 12.5 ~mb \cite{RS84}.

The first measurement of the stellar $^{62}$Ni$(n,\gamma)$ cross section was carried out in 1975 via the time-of-flight technique (TOF) and yielded $<$$\sigma$$>$$_{30keV}$=26.8$\pm$5.0~mb \cite{BS75}. Depending on how the DC component was calculated, the results could change to values between 12.5~mb \cite{RS84} and 35.5~mb \cite{GW84}. The lower cross section is derived if a subthres\-hold resonance is subtracted at thermal energies before the DC cross section is extrapolated with $s$-wave behavior into the stellar keV-region. 

The $^{62}$Ni$(n,\gamma)$ cross section has been measured recently with both, the TOF method \cite{AT05,AAV08} and with the activation method plus AMS \cite{HN05,SW08}. The results from both AMS measurements are in good agreement and also consistent with the recent TOF measurement by \cite{AAV08}. Purely theoretical predictions for the $^{62}$Ni capture cross section at $kT$=30~keV range between 9.7~mb and 21.2~mb, below a weighted average of 26$\pm$2~mb calculated from Refs. \cite{BS75,HN05,AAV08,SW08}.


For the $^{63}$Ni$(n,\gamma)$ cross section experimental data existed neither for the neutron capture channel nor the inverse photodissociation channel. Our measurement of the $^{64}$Ni$(\gamma,n)$$^{63}$Ni can be used to test the theoretical model predictions for this cross section and thus help to improve the predictions for the $^{63}$Ni$(n,\gamma)$ channel.

Stellar model calculations \cite{MP08,SW08} showed the influence of a lower $^{63}$Ni$(n,\gamma)$ cross section during the two phases of the weak $s$ process in massive stars. During core He burning (T$\approx$300~MK) neutron densities are so low that a change in the $^{63}$Ni cross section has no influence because $^{63}$Ni can $\beta$-decay to $^{63}$Cu. However, at the high neutron densities during shell C burning with temperatures of $\approx$1~GK a lower cross section can strongly influence the abundances of $^{63}$Cu (+30\%) and $^{64}$Ni (--20\%).

\subsection{The "Stellar $^{62}$Ni Problem"}
The "stellar $^{62}$Ni problem" was postulated when strong overproductions of ($^{61}$Ni and)  $^{62}$Ni were derived from postexplosive production factors of supernova type II explosions in stars with 15~M$_\odot$, 20~M$_\odot$, and 25~M$_\odot$ \cite{TR02}. These stellar models were carried out with a Maxwellian average cross section of $<$$\sigma$$>$$_{30keV}$=12.5~mb and raised the question whether this overproduction is due to uncertainties in the stellar models or in the nuclear input. An increase in the cross section would lead to more destruction of $^{62}$Ni, but in turn to a increase of the abundances for isotopes with 63$>$$A$$>$90. With the new $^{62}$Ni$(n,\gamma)$ cross section the abundance of $^{62}$Ni goes down by 20\%, but the abundances of all following isotopes up to the next shell-closure at $N$=50 are increased by $\approx$20\%, thus relaxing the "$^{62}$Ni problem".

On the other hand the production of $^{62}$Ni in massive stars could be larger than previously calculated. This possibility might be supported by a recent publication of Marhas et al. \cite{MAG08}, who measured the Ni isotopes in SiC grains with nanoSIMS and found large excesses ($\delta$~$^{61}$Ni/$^{58}$Ni(average)= 622$\permil$ and $\delta$~$^{62}$Ni/$^{58}$Ni(average)= 206$\permil$) of $^{61,62}$Ni in the so-called "X grains" (which are thought to originate from the ejecta of SN type II). These isotopic signatures can only be explained in stellar model calculations by mixing contributions from different zones in the SN II.

\section{Conclusion}
The "stellar $^{62}$Ni problem" has been solved from the nuclear physics side, and a weighted average of $<$$\sigma$$>$$_{30keV}$=26$\pm$2~mb from \cite{BS75,AAV08,HN05,SW08} can be recommended for future stellar modeling. However, the calculated overproduction of $^{62}$Ni is not removed completely by this new determination of the cross section. The remaining overproduction might be due to either a really existing overproduction (possibly supported by $^{61,62}$Ni excesses in presolar X grains) or to problems in stellar modelling.
We have also measured the $^{64}$Ni$(\gamma,n)$ cross section for the first time, leading to the preliminary conclusion that this cross section might be smaller by a factor of two compared to NON-SMOKER predictions.

\section{acknowledments}
We thank the operator teams at the Van de Graaff accelerator at Forschungszentrum Karlsruhe, the ELBE accelerator at Forschungszentrum Rossendorf, and the Tandem accelerator in Garching for their help and support during the measurements.
This research was supported by the DFG cluster of excellence "Origin and Structure of the Universe" (www.universe-cluster.de).

\end{document}